\documentclass[twocolumn,preprintnumbers,amsmath,amssymb]{revtex4}

\usepackage{graphicx}
\usepackage{dcolumn}
\usepackage{bm}

\begin{document}


\title{Twitter-based analysis of the dynamics of collective
attention to political parties}

\author{Young-Ho Eom}
\email{youngho.eom@imtlucca.it}
 \affiliation{IMT Institute for Advanced Studies, Piazza San Francesco 19, 55100 Lucca, Italy}
\author{Michelangelo Puliga}%
 \affiliation{IMT Institute for Advanced Studies, Piazza San Francesco 19, 55100 Lucca, Italy}
\author{Jasmina Smailovi\'{c}}
\affiliation{Jo\v{z}ef Stefan Institute, Jamova 39, 1000
Ljubljana, Slovenia}
\author{Igor Mozeti\v{c}}
\affiliation{Jo\v{z}ef Stefan Institute, Jamova 39, 1000
Ljubljana, Slovenia}
\author{Guido Caldarelli}
\affiliation{IMT Institute for Advanced Studies, Piazza San
Francesco 19, 55100 Lucca, Italy} \affiliation{Istituto dei
Sistemi Complessi (ISC), via dei Taurini 19, 00185 Roma, Italy}
\affiliation{London Institute for Mathematical Sciences, 35a South
Street Mayfair, London, W1K 2XF, UK} \affiliation{Linkalab,
Complex Systems Computational Laboratory, Cagliari, Italy}


\begin{abstract}
Large-scale data from social media have a significant potential to
describe complex phenomena in real world and to anticipate
collective behaviors such as information spreading and social
trends. One specific case of study is represented by the
collective attention to the action of political parties. Not
surprisingly, researchers and stakeholders tried to correlate
parties' presence on social media with their performances in
elections. Despite the many efforts, results are still
inconclusive since this kind of data is often very noisy and
significant signals could be covered by (largely unknown)
statistical fluctuations. In this paper we consider the number of
tweets (tweet volume) of a party as a proxy of collective
attention to the party, identify the dynamics of the volume, and
show that this quantity has some information on the elections
outcome. We find that the distribution of the tweet volume for
each party follows a log-normal distribution with a positive
autocorrelation of the volume over short terms, which indicates
the volume has large fluctuations of the log-normal distribution
yet with a short-term tendency. Furthermore, by measuring the
ratio of two consecutive daily tweet volumes, we find that the
evolution of the daily volume of a party can be described by means
of a geometric Brownian motion (i.e., the logarithm of the volume
moves randomly with a trend). Finally, we determine the optimal
period of averaging tweet volume for reducing fluctuations and
extracting short-term tendencies. We conclude that the tweet
volume is a good indicator of parties' success in the elections
when considered over an optimal time window. Our study identifies
the statistical nature of collective attention to political issues
and sheds light on how to model the dynamics of collective
attention in social media.
\end{abstract}

\maketitle

\section*{Introduction}
As social animals, since a long time ago, humans have
communicated, exchanged opinions, and tried to reconcile their
conflicts by means of social instruments. Despite their recent
introduction, social media and web-based services such as Google,
Twitter, Facebook, and Wikipedia have already dramatically changed
the way in which people make relationships, interact with others,
and acquire information. Differently from the past, such
activities help people to overcome the physical and geographical
limitations of human interactions.

When people use social media and web services, a huge amount of
digital ``footprints'' (i.e., data) are created and simultaneously
recorded. These ``footprints'' can provide us novel opportunities
to observe collective behaviors at unprecedented scales. For this
reason, the data are generally regarded as crucial instruments in
order to understand the complex and collective behaviors in our
social and technological
systems~\cite{Lazer2009,Giles2012,Vespignani2009,Conte2012,Moat2014}.
Despite the recent appearance of these computer-based social
media, there is already a large number of studies describing and
forecasting collective behaviors emerging from them. For example,
large scale network analysis based on Twitter and Facebook data
have revealed the structure of social networks of tens of millions
of people~\cite{Kwak2010,Ugander2011}. Twitter data have been used
to identify spreading patterns of popular
information~\cite{Lerman2010,Cheng2014}, classes of dynamical
collective attention~\cite{Lehmann2012}, linguistic usage patterns
on worldwide scale~\cite{Mocanu2013}, and political
activity~\cite{Conover2011,Conover2012,Conover2013}. From Facebook
data it has been possible to distinguish difference in consumption
patterns between science and conspiracy
information~\cite{Bessi2015}. Further cross-cultural differences
in evaluation of historical figures were identified based on
multilingual Wikipedia data~\cite{Eom2013,Eom2015}, and social
media usage patterns are used to find out unemployment in local
regions~\cite{Llorente2014}. Finally, users' query logs on search
engines help to anticipate the spreading of
flu~\cite{Ginsberg2009} or dynamics of stock
market~\cite{Bordino2012,Curme2014}, and Wikipedia activity data
was used to predict movies' box office~\cite{Mestyan2013}.

Predictions of elections based on social media data have various
advantages with respect to other methods (such as traditional
opinion polls). Firstly, we deal with large scale samples,
secondly, the flow of data is such that we can get real time
responses, and finally, we have low costs of data collection. For
these reasons, social media data received (and probably will
receive even more in the future) a great attention by
practitioners and scientists. The key question will be whether
relevant information on elections can be extracted from social
media data or not. It is now known that in certain cases we can
have indications on elections results, but the degree of
reliability of this method has to be
improved~\cite{GayoAvello2013}. For example, both positive
~\cite{Borondo2012,DiGrazia2013,Caldarelli2014,Smailovic2014} and
null relations~\cite{GayoAvello2012,Jungherr2013} between social
media activity and election outcomes have been observed so far. In
order to improve this method of forecast, some scientists suggest
to complement tweet volume analysis with sentiment analysis of
tweets, i.e., identification of positive or negative
sentiment~\cite{OConnor2010,Smailovic2014}. Nevertheless, reliable
methods of sentiment analysis for political tweets are still
lacking~\cite{Yasseri2014}. Intuitively, mentions of political
parties or politicians in social media can be considered as
expressions of people's attention to them. However, there is no
guarantee that all of the mentions in social media correspond to
the supports for the parties in elections. People post tweets on
political parties and politicians for various reasons, such as
expressions of support, disappointment, or sarcasm. In other
words, dynamics of tweet activity can be driven not only by
popularity of parties or politicians but also by other reasons.
Therefore it is necessary to understand dynamics of collective
attention to political parties or politicians in social media,
since such understanding will be a cornerstone to separate the
``signal'' from the ``noise'' in the dynamics of collective
attention in social media.

In this paper we consider tweet volumes about political parties as
proxies of collective attention to the parties and by
investigating the dynamics of tweet volumes we try to assess their
relation (and forecasting power) with the final results of
elections. For such purposes, we identify dynamical and
statistical characteristics of daily tweet volumes of political
parties during election periods. We find that the distributions of
daily tweet volume of each political party is in good agreement
with log-normal distribution\cite{Mitzenmacher2004}. This
observation indicates that the average behavior of daily tweet
volume may have some information, yet large fluctuations can be
behind the average. Thus the prediction based on too short-term
Twitter data may not be consistent. On the other hand, we observed
positive autocorrelation of daily tweet volume of each party in
short term. This means the time series of daily tweet volume
largely depends on the previous activity (i.e., the existence of
short-term tendency). Thus, averaging over too long-term periods
can destroy the signal. We also measure that the distribution of
the logarithmic ratio of two consecutive daily tweet volumes for
each party follows a normal distribution and the ratio is
independent of time. These two observations allow us to describe
properly the dynamics of daily tweet volume as a geometric
Brownian motion\cite{Marathe2005}. In the end, we checked whether
there is an optimal period of averaging tweet volumes which not
only reduce the fluctuation but also keep the short term tendency
of tweet volumes. Our analysis suggests what really tweet volume
of each political party means in a quantitative way and sheds
light on how we can separate the noise and the signal for better
prediction using social media data.

\section*{Materials and Methods}
\subsection*{Data description}

In this paper, we consider data collected on Twitter
(twitter.com), a microblogging platform used by millions of
bloggers. In Twitter, each user can freely post short messages (up
to $140$ characters) called ``tweets" to its followers. Twitter
provides application programming interfaces (APIs) to access
tweets and information about tweets and users. The potential bias
of Twitter APIs was discussed by a recent
research~\cite{Gonzalez-Bailon2014}. We mainly consider daily
tweet volume $V_p(t)$ of a given political party $p$ at day $t$.
To identify dynamics of daily tweet volume of political parties in
Twitter, we consider three elections in two European countries:
\emph{European Parliament election of 2014 in Italy (Euro14)},
\emph{Italian general election of 2013 (Italy13)}, and
\emph{Bulgarian general election of 2013 (Bulgaria13)}. By using
Twitter API, we collected general tweets around election days and
then considered only tweets posted in local languages (i.e.,
Italian or Bulgarian) from the starting day of data collection to
the day before the election day. We used the implemented automatic
language detection system of Twitter to identify the language of
tweets. For the Bulgarian case, the Twitter language detection
mechanism often did not distinguish between Bulgarian and
Macedonian, which are very similar. We therefore implemented our
own language detection, based on a Bayesian classifier, trained on
a large corpus of over five million words for each language. Here
one day is defined as a time window from 00:00:00 to 23:59:59 of
the day in local time for the Italian cases and Greenwich Mean
Time for the Bulgarian case. For the cases of election in Italy
(i.e., \emph{Euro14} and \emph{Italy13}), we define the number of
tweets $V_p(t)$ for a given political party $p$ as the number of
tweets mentioning the leaders' names (only family names) of
political parties $p$ or the leaders' twitter accounts at the day
$t$. This is because, in Italian cases, the names of leaders are
widely used to represent the political
parties~\cite{Caldarelli2014}. The overview summary of three data
sets are represented in Table~\ref{table1}.

\begin{itemize}
\item \emph{European Parliament election of 2014 in Italy
(Euro14)}: We collected 12,535,469 tweets posted between 21 April
2014 and 12 June 2014 in total. Of this sample, we extracted
3,413,214 Italian tweets between 22 April 2014 and 23 May 2014.
The election day was 24 May 2014~\cite{Euro2014Result}.

\item \emph{Italian general election of 2013 (Italy13)}: We
collected 7,755,063 tweets posted between 11 November 2012 to 3
March 2013 in total. Of this sample, we extracted 3,796,754
Italian tweets from 1 January 2013 to 22 February 2013. The
election days were 23 and 24 February 2013~\cite{Italy2013Result}.

\item \emph{Bulgarian general election of 2013 (Bulgaria13)}: The
raw tweet data is based on collected 16,077 tweets posted between
29 April 2013 to 27 May 2013 in total~\cite{Smailovic2014}. Out of
this sample, we extracted 5,817 tweets from 29 April to 11 May
2013. The election day was 12 May 2013~\cite{Bulgaria2013 Result}.
In this case we consider both, the names of political parties and
the names of their leaders. The retrieval of the Bulgarian tweets
was performed by the Gama System company
(http://www.gama-system.si/en/) and their Gama System$^\circledR$
PerceptionAnalytics platform (http://demo.perceptionanalytics.net)
\end{itemize}

Detailed information on each party in each election is given in
Table~\ref{table2}.

\subsection*{Geometric Brownian motion}
Defining a geometric Brownian motion for the daily tweet volume
$V_p(t)$ (for a party $p$) means that $V_p(t)$ satisfies the
following stochastic differential
equations~\cite{Wilk1968,Marathe2005}:
\begin{equation}\label{eq:GBMeq}
dV_p(t)=\mu V_p(t)dt+\sigma V_p(t)dW_t
\end{equation}
where $W_t$ is Wiener process or Brownian motion, and $\mu$ and
$\sigma$ are constants. In particular, $\mu$ represents the
``drift'' (i.e., trend) and $\sigma$ represents the ``volatility''
(i.e., random noise) of $V_p(t)$. Eq.~\ref{eq:GBMeq} has an
analytic solution under It\={o}'s interpretation~\cite{Ross1998}
as following:
\begin{equation}\label{eq:GBM}
V_p(t)= V_p(0)exp((\mu-\frac{\sigma^2}{2})t+\sigma W_t)
\end{equation}
where $V_p(0)$ is the initial value.

Taking logarithm of both sides of Eq.~\ref{eq:GBM}, we get:
\begin{equation}\label{eq:LogGBM}
log(V_p(t))= log(V_p(0))+(\mu-\frac{\sigma^2}{2})t+\sigma W_t
\end{equation}

Since $\langle W(t) \rangle=0$, the expectation value of
$log(V_p(t))$ is given in the following equation:
\begin{equation}\label{eq:AveLogGBM}
\langle log(V_p(t))\rangle = log(V_p(0))+(\mu-\frac{\sigma^2}{2})t
\end{equation}

\section*{Results}
The main results of this paper are summarized as follows.
(i) We find that the daily tweet volumes of political parties
before elections follow log-normal distributions and have positive
autocorrelations over short terms.
(ii) The daily volume evolution can be described by means of
geometric Brownian motion.
(iii) If we want to consider the average behavior of daily tweet
volume, it is necessary to consider long enough period for
reducing statistical fluctuations, but not too long, to not
destroy short-term memories with relevant information.

\subsection*{Indication from tweet volumes}
We consider dynamics of daily tweet volumes of political parties
in three elections (\emph{Euro14}, \emph{Italy13}, and
\emph{Bulgaria13}) based on the Twitter data collected as
described in the Method section. The time series of daily tweet
volume $V_p(t)$ of a political party $p$, before and after each
election day, are represented in Fig.~\ref{fig1}.
Sharp peaks of daily tweet volumes of parties on the election days
and on the day after election days suggest the daily tweet volumes
reflect the attentions of the public to the elections. On the
other hand, other notable peaks are also observed much earlier
than the election days, which indicate the daily tweet volumes may
be activated by other reasons than election issues, such as
scandals of politicians, their appearances in the press or mass
media, or other political activities~\cite{Gonzalez-Bailon2013}.

For these three election cases, we want to check if we can get an
indication on the election outcomes simply considering daily tweet
volume of parties or its simple functions as reported in some
studies~\cite{Borondo2012,DiGrazia2013,Caldarelli2014}. As shown
in Fig.~\ref{fig1}, the daily tweet volume for each party shows
different prediction power for election outcomes depending on
elections. The ordering of parties in Fig.~\ref{fig1} is
determined by actual rankings based on number of votes in the
elections (See Table~\ref{table2}). In the case of
\emph{Bulgaria13} (Fig.~\ref{fig1}(C)), during the whole
observation period, rankings by the daily tweet volumes are the
same as the actual election outcome. In the case of \emph{Euro14}
(Fig.~\ref{fig1}(A)), for most of observation days, daily tweet
volume predicted well the election outcome. In the \emph{Italy13}
case (Fig.~\ref{fig1}(B)), the prediction is less effective than
the other two cases especially in early days. In the
\emph{Italy13} case, the  rankings predicted by analysis change
frequently with the day, therefore making the forecast not very
reliable. However, we cannot conclude that this is a failure of
the method, since it could actually reflect the real dynamics of
voters' opinions. Indeed, according to the opinion polls in
Italy~\cite{OpinionPollItaly2013}, M5S had low support from the
public in the early period of the campaign. Also it is notable
that the \emph{Italy13} case is a typical `too close to call' case
(See Table~\ref{table2} for the actual number of votes) to
evaluate the prediction power.

\subsection*{Description of fluctuations in tweet volumes}
The observed fluctuations in daily tweet volumes can distort not
only prediction of parties' rankings in elections but also the
prediction on parties actual  votes in the elections. While it
seems possible to forecast rankings in some elections there is
still some work to be done to anticipate the number of actual
votes. Indeed, depending on the observation period, the prediction
of the number of votes varied because strong fluctuations exist in
daily tweets volumes for each party. Similar behaviors were also
observed previously~\cite{Borondo2012,Caldarelli2014}.

If the daily tweet volumes of parties show strong fluctuations, it
is necessary at least to describe the statistical patterns of the
evolution of this quantity. To this aim, we consider distributions
of daily tweet volume $V_p$ for the given time interval from the
initial day of data collection to the day before the elections.
From visual inspection, this quantity seem to follow a fat-tailed
like ``log-normal'' distributions (Fig.~\ref{fig2}(A), (C), and
(E)). Due to the small number of data samples, we represented the
cumulative distribution functions. To determine whether the daily
tweet volumes follows or not log-normal, we consider Q-Q plot
(quantile-quantile plot)~\cite{Wilk1968} of logarithm of $V_p$ as
shown in Fig.~\ref{fig2}(B), (D), and (F).

Note that if the points in the Q-Q plot are close to $y=x$ line,
the data is more likely to follow the theoretical distribution
(i.e., normal distribution in this case). As shown in
Fig.~\ref{fig2}(B), (D), and (F), in most of the cases we can
conclude that the daily tweet volumes follow log-normal
distributions since logarithms of the volumes follow normal
distributions as shown in the Q-Q plots. Such fat-tailed shape
means that even if the daily tweet volume may provide relevant
information on the dynamics of collective attention to political
issues, this information can be largely hidden by statistical
fluctuations. Thus, in spite of some prediction power, it is not
easy to predict the election outcome very accurately beyond the
rankings due to the fluctuations.

We then checked whether the dynamics of the daily tweet volumes
$V_p$ can be described by a constant volume with fluctuations, or
if there exist higher orders in the dynamics. First, in order to
check if the daily tweet volumes can be described as a constant
volume term with a noise volume term, we consider autocorrelation
$R_p$ of the daily tweet volume $V_p(t)$ for each party $p$. If we
can consider $V_p(t) = V_0 + E_t$, where $V_0$ is a constant and
$E_t$ is an error term, $V_p(t)$ will move around $V_0$ as a
random signal without any short or long term tendency. In this
case, autocorrelation of $V_p$ will be zero. The autocorrelation
measures how similar is the original time series of a variable to
the lagged time series of the variable. We can measure
autocorrelation $R_{p}(\tau)$ of daily tweet volume for a party
$p$ with a lagged time $\tau$ by the Pearson's coefficient between
original tweet volume from day $t=0$ to $t=t_e-1-\tau$ and the
same tweet volume from day $t=\tau$ to $t=t_e-1$ for a given party
$p$ and $\tau$:
\begin{equation}\label{eq:AC}
R_{p}(\tau) =
\frac{1}{t_e-\tau}\sum_{t=0}^{t_e-1-\tau}\frac{(V_p(t)-\langle V
\rangle)(V_p(t+\tau)-\langle V' \rangle)}{\sigma_p\sigma'_p}
\end{equation}

Here, $\langle V \rangle$ ($\langle V' \rangle$) is the average
daily tweet volume for party $p$ from day $t=0$ ($t=\tau$) to day
$t=t_e-1-\tau$ ($t=t_e-1$), $\sigma_p$ ($\sigma'_p$) is the
standard deviation, and $t_e$ is the election day. Thus
$R_{p}(\tau)$ quantifies the correlation between original time
series of daily tweet volume $V_p(t)$ with $\tau$ day-lagged time
series $V_p(t+\tau)$ of original daily tweet volume. If
$R_{p}(\tau)=1$, the time series has strongly increasing or
decreasing tendency with period of $\tau$. If $R_{p}(\tau)=-1$,
the time series shows `up and down' or zigzag pattern with period
of $\tau$. If $R_{p}(1) \approx 0.0$, then we can consider
$V_p(t)$ such that $V_p(t)=V_0 + E_t$ where $V_0$ is a constant
and $E_t$ is an error (or noise) term as described above. As shown
in Fig.~\ref{fig3}, we observed positive autocorrelations
$R_{p}(1) \geq 0.2$ for all of the cases. This means the daily
tweet volume for parties have some `increasing' or `decreasing'
patterns for some time intervals and cannot be described by a
simple constant plus error model. However, $R_{p}(\tau\geq 2)
\approx 0$ in some cases. In these cases the tendency do not last
long. While $R_{p}(\tau\geq 2) \geq 0.4$ for M5S and AET in
\emph{Euro14} (Fig.~\ref{fig3}(A)), for M5S in \emph{Italy13}
(Fig.~\ref{fig3}(B)), and for DPS and ATAKA in \emph{Bulgaria13}
(Fig.~\ref{fig3}(C)). These cases show more persistent tendency.

\subsection*{A model of fluctuations in tweet volume}
The observed log-normal distributions of daily tweet volumes for
parties suggest that its underlying dynamics can be described by a
geometric Brownian motion (GBM)~\cite{Wilk1968}. This means that
the logarithm of the variable follows a Brownian motion with a
drift, a situation that  often describes the dynamics of company
prices in stock markets~\cite{Marathe2005}.

To verify this assumption we need to check if the logarithmic
ratio $r_p(t)=log(V_p(t+1)/V_p(t))$ follows a normal distribution
and if the same ratio is independent of
time~\cite{Ross1998,Marathe2005}.

Regarding the first point, we show  in  Fig.~\ref{fig4}(A), (C),
and (E) the cumulative distribution functions of $r$ for every
party. To confirm that they are indeed normally distributed, we
consider the Q-Q plots for each party as shown in
Fig.~\ref{fig4}(B), (D), and (F) (as described in
Fig.~\ref{fig2}). The Q-Q plots strongly support the normality of
the logarithmic ratio $r_p(t)$ (the points approximately lie on
$y=x$ line). As for the second point we consider the scatter plots
of the logarithmic ratio $r_p(t)=log(V_p(t+1)/V_p(t))$ as shown in
Fig.~\ref{fig5}. From Fig.~\ref{fig5} we can see that the ratio
$r_p(t)$ for every party is independent of time $t$.

By fulfilling the above hypotheses, we can consider
Eq.~\ref{eq:AveLogGBM} as a GBM model for dynamics of $V_p(t)$. By
linear fitting of the data with Eq.~\ref{eq:AveLogGBM}, we can
determine the value of $\mu-\frac{\sigma^2}{2}$ and $log(V_p(0))$.
Then we get the value of $\sigma$ from the fluctuations between
the data and the GBM model. The obtained values of $\mu$,
$\sigma$, and $V_0$ are represented in Table~\ref{table3}.

Fig.~\ref{fig6} shows the dynamics of $V_p(t)$ for each party  $p$
(red lines) and the corresponding GBM model
$V_p(0)exp((\mu-\frac{\sigma^2}{2})t)$ (blue dashed lines). As
guidelines, GBM$+\sigma$ model
$V_p(0)exp((\mu-\frac{\sigma^2}{2})t+\sigma)$ (green dashed lines)
and GBM$-\sigma$ model
$V_p(0)exp((\mu-\frac{\sigma^2}{2})t-\sigma)$ (cyan dashed lines)
are also represented in Fig.~\ref{fig6}. Indeed, the GBM model
describes well the dynamics of daily tweet volume in the data as
shown in Fig.~\ref{fig6} although there are some large spikes,
which are beyond the GBM$+\sigma$ model, in the dynamics. Also the
obtained values of $\mu$ and $\sigma$ explain the observed strong
autocorrelations of daily tweet volumes. For example, M5S in
\emph{Euro14} and \emph{Italy13} has relatively high $\mu$ but low
$\sigma$, thus the dynamics of daily tweet volume of M5S in
\emph{Euro14} and \emph{Italy13} has relatively strong drift with
weak fluctuations. This leads the dynamics to high
autocorrelations in longer term (i.e., a strong tendency with low
volatility).

\subsection*{Tweet volumes and election outcomes}
Until now we mainly focused on the dynamical properties and the
modelling of daily tweet volumes of political parties in order to
describe the properties of data fluctuations.
Anyhow, the simplest way of reducing fluctuations will be
averaging out (or cumulating) the daily tweet volumes. However,
positive autocorrelation and short-term memory of the volumes
imply that if we consider too long time interval for averaging, we
might lose short term increasing or decreasing tendency in the
dynamics. In other words, if we consider too long period, the
recent relevant signals from tweet volumes can be hidden by old
tweet volumes. In addition, if we consider tweet volume in days
much earlier than the election day, other types of `noise'
compromise the `signal'. Twitter users typically do not pay much
attention to elections before the campaign actually starts, even
though they may mention ``politics" in their tweets. Thus it is
necessary to find out how long time interval has to be considered
to get optimal results in practical sense.

To identify the optimal time interval of averaging daily tweet
volume of a given political party, we consider the tweet volume
$\bar{V_p}(\lambda)$ of a party $p$ averaged from the day before
the election to the $|\lambda|$ days before as follows:

\begin{equation}\label{eq:average}
\bar{V_p}(\lambda) =
\frac{1}{|\lambda|}\sum_{t=t_e-|\lambda|}^{t_e-1}V_p(t).
\end{equation}
Here $t_e$ is the election day, $\lambda$ is a negative integer,
and $|\lambda|$ is the absolute value of $\lambda$ that represents
the number of days to wait for the election day (i.e.,
$\lambda=-2$ means two days before the election day).

Fig.~\ref{fig7} shows the rankings of parties ordered by
$\bar{V}_p(\lambda)$ for each time interval from the day before
the election day to the $|\lambda|$ days before the election. For
the case of \emph{Euro14} (Fig.~\ref{fig7}(A)), until
$\lambda=-14$, we can get the accurate prediction. For the case of
\emph{Italy13}, the optimal length of time interval for accurate
prediction will be from $\lambda=-2$ to $\lambda=-11$. Indeed, M5S
performed much better than the expectation before the election and
the support for M5S was rapidly growing during the campaign. This
pattern is vividly reflected in Fig.~\ref{fig7}(B). If we consider
$\lambda=-14$, then the prediction based tweet volume M5S
anticipated M5S will be the third thanks to the low supports for
M5S in earlier period of the campaign. On the other hand, all
considered $\lambda$ show accurate and consistent prediction in
the case of \emph{Bulgaria13} (Fig.~\ref{fig7}(C)), as expected
from Fig.~\ref{fig1}.


\section*{Discussion}
Social media permeate all levels of society rapidly and widely. A
huge amount of data on collective behaviors are being generated
from these social media. This phenomenon promotes quantitative
analysis of these data, with the goal to understand collective
behaviors and predict them in effective and efficient ways. In
this paper, we analyzed dynamics of daily tweet volumes of
political parties on Twitter, when approaching elections,
identified statistical patterns of the daily tweet volumes of
parties, and described the dynamics of volume with geometric
Brownian motion (GBM). We found that the daily tweet volume of a
given political party follows a broad distribution like
log-normal, and has positive autocorrelation over a short time
period. Finally, we identified there is an optimal period of
averaging tweet volumes which not only reduce the fluctuation but
also keep the short term tendency of tweet volumes. Our analysis
shows that daily tweet volumes could have a limited prediction
ability of election outcomes and that this limitation is caused by
their strong fluctuations.

\begin{figure*}[h]
\includegraphics[width=170mm, angle=0]{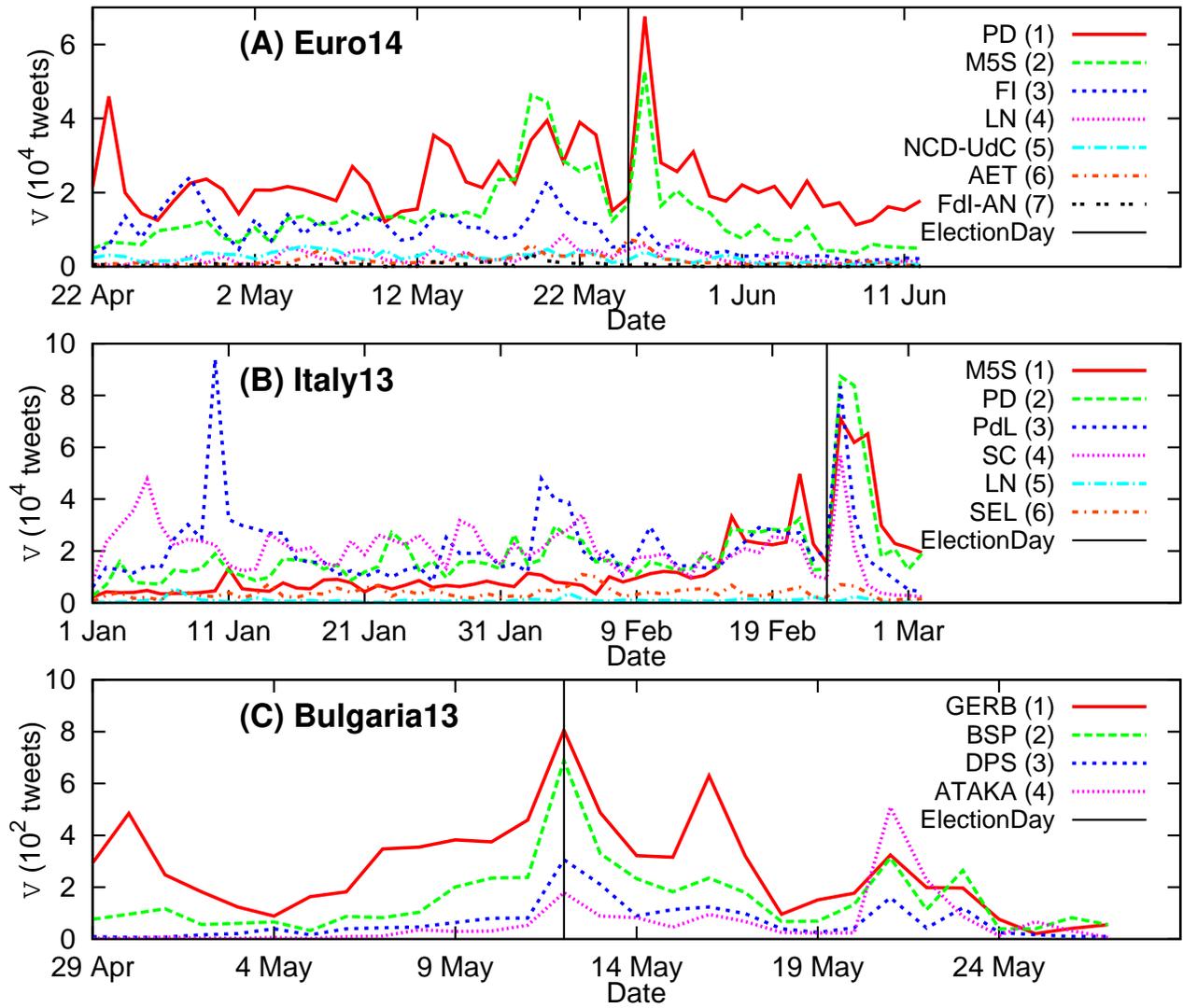}
\caption{{\bf Daily tweet volume for each party around elections.
} The ordering of parties (i.e., the numbers in parentheses) is
based on actual ranking in the election. (A) \emph{Euro14}. 1st:
PD. 2nd: M5S. 3rd: FI. 4th: LN. 5th: NCD-UdC. 6th: AET. 7th:
FdI-AN. (B) \emph{Italy13}. 1st: M5S. 2nd: PD. 3rd: PdL. 4th: SC.
5th: LN. 6th: SEL. (C) \emph{Bulgaria13}. 1st: GERB. 2nd: BSP.
3rd: DPS. 4th: ATAKA.}. \label{fig1}
\end{figure*}

\begin{figure*}[h]
\includegraphics[width=170mm, angle=0]{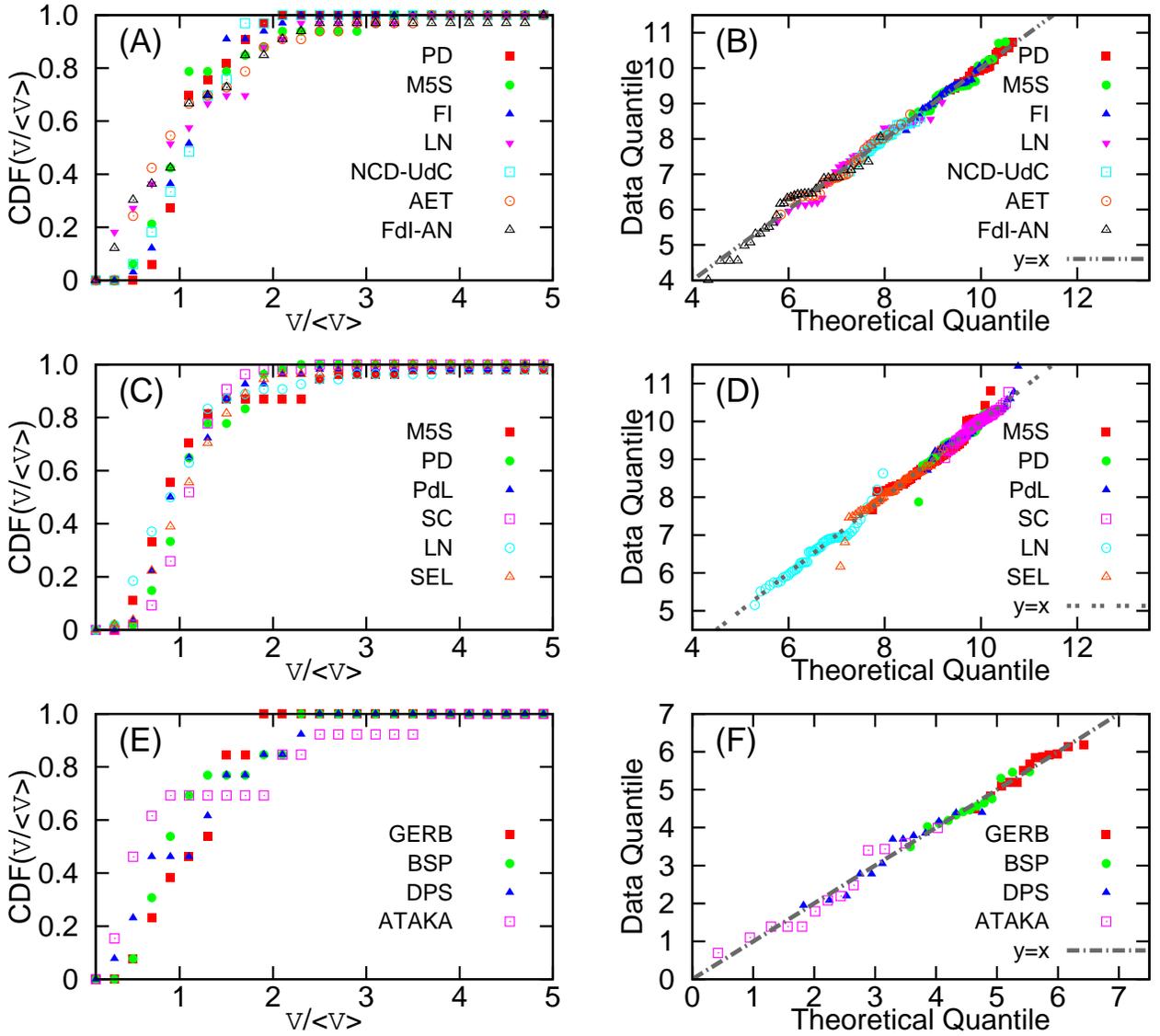}
\caption{{\bf Cumulative distribution functions (CDF) of daily
tweet volumes (A, C, E) and Q-Q plots of logarithms of daily tweet
volumes for each political party (B, D, F).} Each volume in CDF is
normalized by the average $\langle V \rangle$. (A) CDF of daily
tweet volume of \emph{Euro14}. (B) Q-Q plot of \emph{Euro14}. (C)
CDF of daily tweet volume of \emph{Italy13}. (D) Q-Q plot of
\emph{Italy13}. (E) CDF of daily tweet volume in
\emph{Bulgaria13}. (f) Q-Q plot in \emph{Bulgaria13}. Note that
Q-Q plot is for logarithm of daily tweet volume. Theoretical
quantile in the Q-Q plot is based on normal distribution. Thus if
the points in the Q-Q plot lie on $y=x$ line, the daily tweet
volume follows a log-normal distribution since the logarithm of
the volume follow a normal distribution.} \label{fig2}
\end{figure*}

\begin{figure*}[h]
\includegraphics[width=170mm, angle=0]{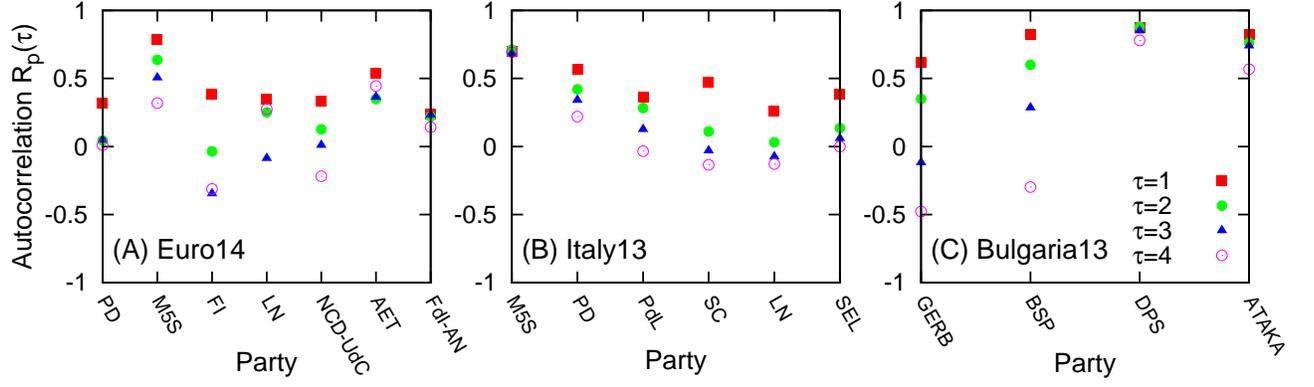}
\caption{{\bf Autocorrelation of daily tweet volume for each
political party.}  Autocorrelation coefficient $R_{p}(\tau)$ is
given by $R_{p}(\tau) =
\frac{1}{t_e-\tau}\sum_{t=0}^{t_e-1-\tau}\frac{(V_p(t)-\langle V
\rangle)(V_p(t+\tau)-\langle V' \rangle)}{\sigma_p\sigma'_p}$.
Here $\langle V \rangle$ ($\langle V' \rangle$) is the average
daily tweet volume for party $p$ from day $t=0$ ($t=\tau$) to day
$t=t_e-1-\tau$ ($t=t_e-\tau$), $\sigma_p$ ($\sigma'_p$) is the
standard deviation, and $t_e$ is the election day. (A)
\emph{Euro14}. (B) \emph{Italy13}. (C) \emph{Bulgaria13}.}
\label{fig3}
\end{figure*}

In order to overcome the limited prediction power of the daily
tweet volume, one needs to understand what causes statistical
fluctuations of Twitter activity and to separate the signal from
the noise in tweet volumes. Universal features of fluctuations
with the form of log-normal distributions imply that there might
be a single underlying mechanism for the fluctuations, such as
multiplicative processes~\cite{Mitzenmacher2004}. In particular,
the driving mechanisms of peaked activities, which cause large
fluctuations, should be understood. For instance, Silvio
Berlusconi is a popular figure in Italian politics and society. He
therefore receives a large number of Twitter mentions not only by
his supporters but also by his opponents; often these mentions are
not just about politics but also about his private life. For
example, on 9 Jan. 2013, a sharp peak of FI (i.e., mentioning
Berlusconi) in Fig.~\ref{fig1} was observed. From the news on this
day we concluded that an Italian court fixed the financial
consequences of his divorce and that he was charged with the
accusation of prostitution with a minor (at the time of
publication of this article the trial ended and he was sentenced
not guilty). This example clearly illustrates that the peaks could
stem not only from election issues but also from private issues of
the politicians. This also means that one needs to consider the
roles of mass media for daily tweet volumes of political parties.
All these factors can have significant influence on tweet volumes
of political parties or politicians. Systemic consideration of
these factors can give us some hints about the amount of the
fluctuations originating from the endogenous or exogenous
mechanisms.

\begin{figure*}[h]
\includegraphics[width=170mm, angle=0]{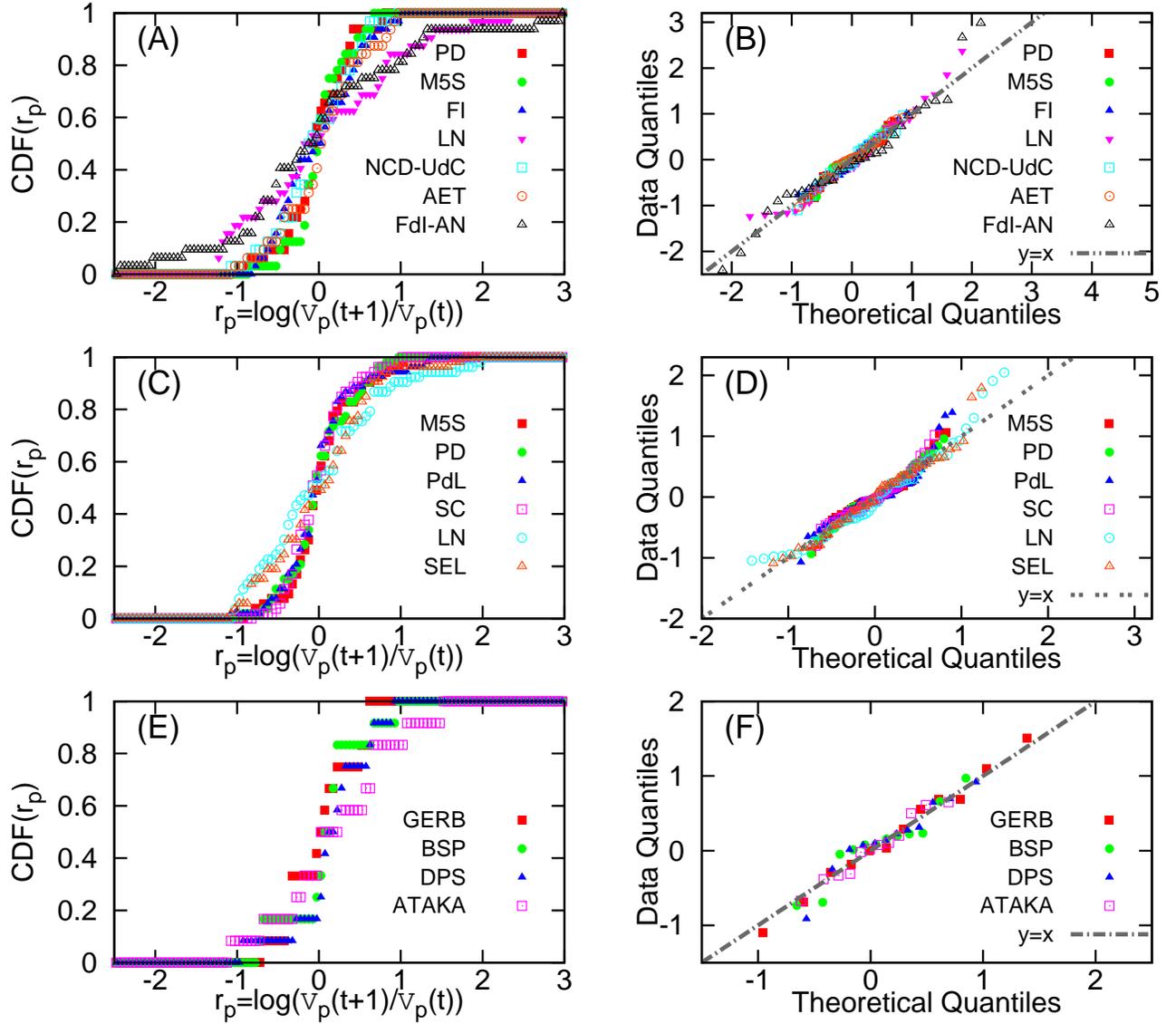}
\caption{{\bf Normality of the logarithmic ratio
$r_p(t)=log(V_p(t+1)/V_p(t))$ of two consecutive tweet volumes of
party $p$}. Cumulative distribution functions of the log ratio for
each party are represented in (A) \emph{Euro14}. (C)
\emph{Italy13}. (E) \emph{Bulgaria13}. The Q-Q plots of the log
ratio $r(t)$ for each party are also represented in (B)
\emph{Euro14}. (D) \emph{Italy13}. (F) \emph{Bulgaria13}. The
theoretical quantile is based on normal distribution. In the Q-Q
plot, if the points lie on $y=x$, it means the log ratio follow a
normal distribution. } \label{fig4}
\end{figure*}

\begin{figure*}[h]
\includegraphics[width=170mm, angle=0]{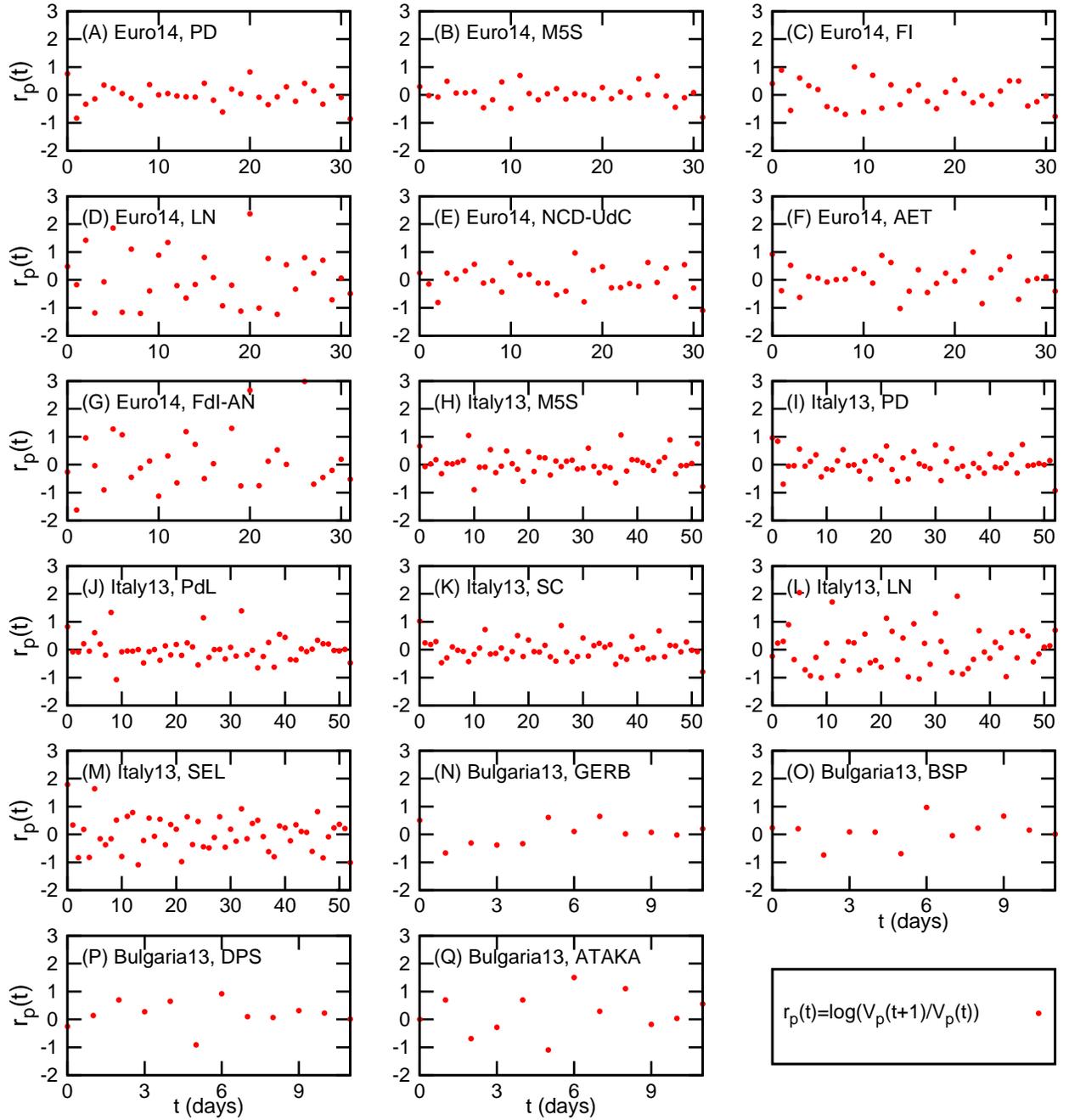}
\caption{{\bf Scatter plot of time $t$ and log ratio
$r_p(t)=log(V_p(t+1)/V_p(t))$ for each party $p$.} Here $V_p(t)$
is the tweet volume of the party $p$ at time $t$.} \label{fig5}
\end{figure*}

\begin{figure*}[h]
\includegraphics[width=170mm, angle=0]{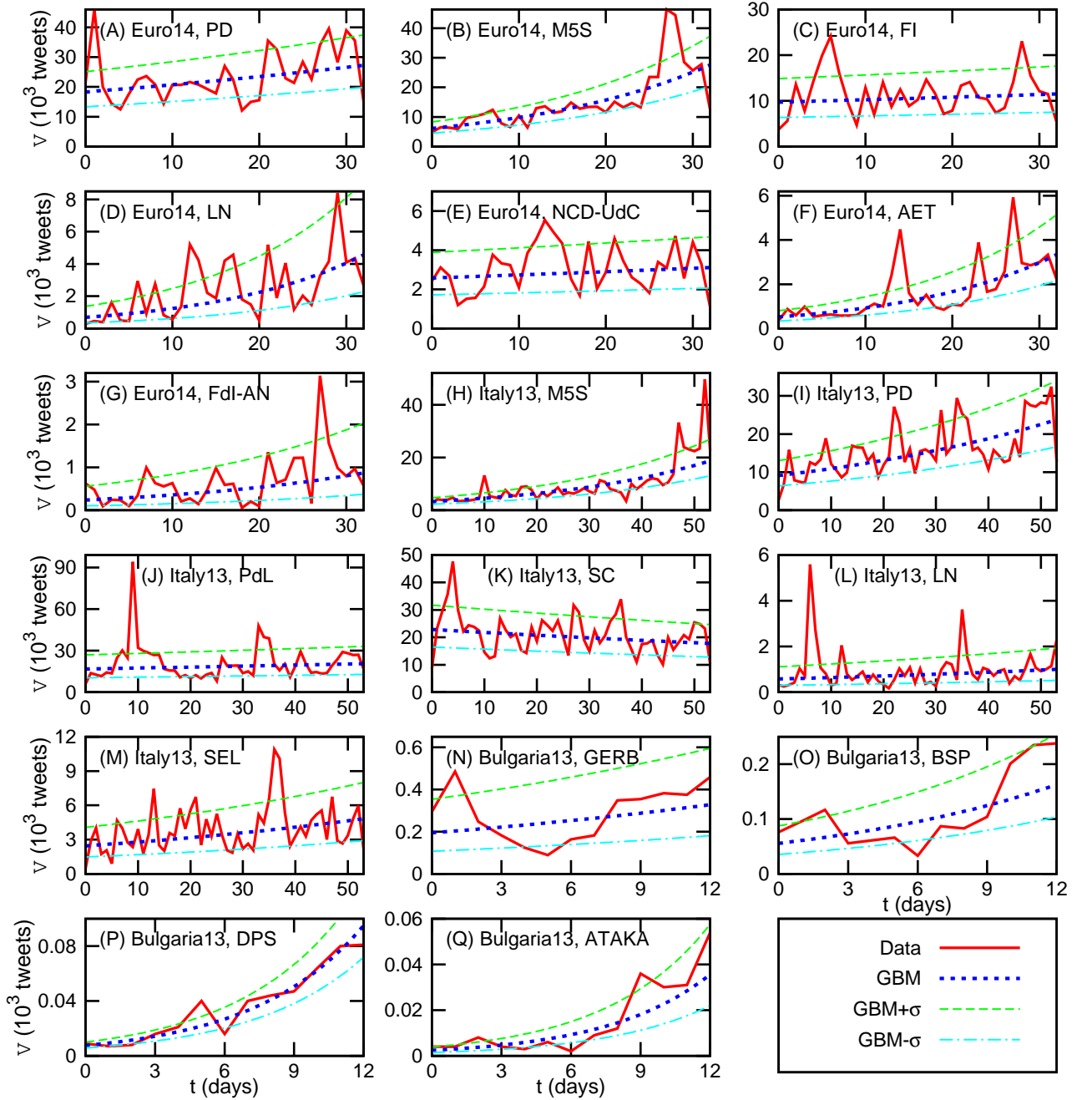}
\caption{{\bf Dynamics of daily tweet volume for each party
represented by data and by the GBM model.} In the GBM model, the
expected volume $V(t)$ at time $t$ is given by $V_p(t)=V_p(0)
exp((\mu-\frac{\sigma^2}{2})t)$. In the GBM$+\sigma$ model,
$V_p(t)=V_p(0) exp((\mu-\frac{\sigma^2}{2})t+\sigma)$ while
$V_p(t)=V_p(0) exp((\mu-\frac{\sigma^2}{2})t-\sigma)$ in the
GBM$-\sigma$ model. The values of parameters $\mu$, $\sigma$, and
$V(0)$ are given in Table~\ref{table3}.} \label{fig6}
\end{figure*}

\begin{figure*}[h]
\includegraphics[width=170mm, angle=0]{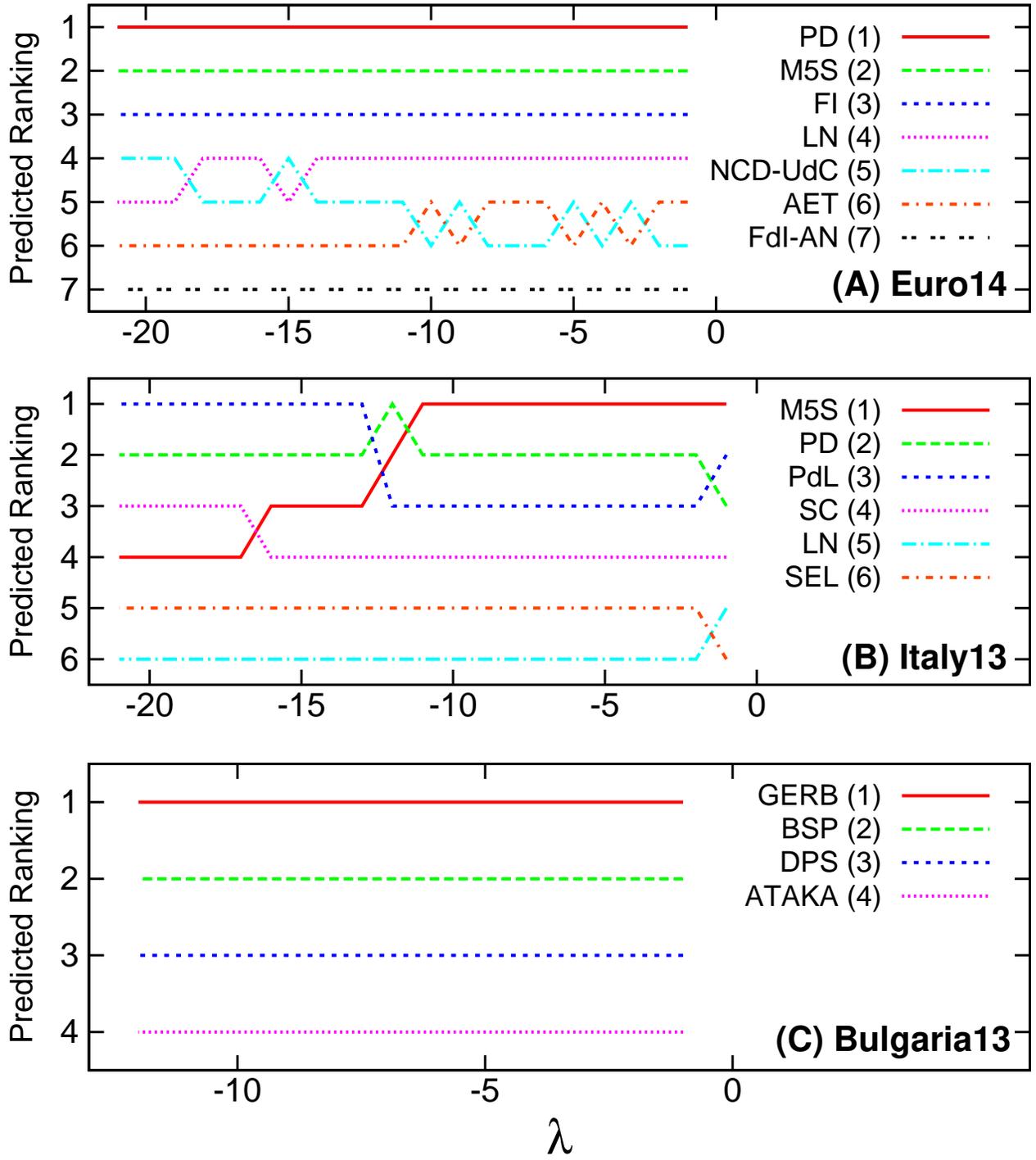}
\caption{{\bf Predicted ranking determined by tweet volume
$\bar{V_p}(\lambda)$ averaged from the day before the election to
the $\tau$ days before the election.} $\bar{V_p}(\lambda)$ is
given by Eq.~\ref{eq:average}. The numbers in parentheses
represent actual rankings of the parties in the election. (A)
\emph{Euro14}. (B) \emph{Italy13}. (C) \emph{Bulgaria13}. }
\label{fig7}
\end{figure*}

Expanding the point of view, it would be interesting to identify
whether the dynamics after the election also can be described as a
GBM or not. If possible, the GBM model for the dynamics after the
election might have different drift ($\mu$) and volatility
($\sigma$) terms in Eq.~\ref{eq:AveLogGBM} from the ones in the
current GBM model for the dynamics before the election. Because,
as shown in Fig. 1, the dynamics of tweet volume typically shows a
peak on the election day or the day after election and show
decreasing patterns hereafter. This implies the drift (i.e.,
tendency) term of the GBM might be changed after the election
since the collective attention was moved to other issues. In order
to describe the dynamics after election as a GBM, it is necessary
to test the normality of logarithmic ratio of consecutive tweet
volume and time-independence of the ratio as done in
Fig.~\ref{fig4} and Fig.~\ref{fig5}. For these tests, we need to
consider tweets data-set collected after the elections.

Not only single social media but also multiple social media can be
considered to predict the election outcome. For instance,
Wikipedia and search engine data have been used to forecast
elections outcomes~\cite{Yasseri2014}, and sentiment analysis was
suggested for reinforcing the forecasting performance. Checking
the validity of combined social media data will be one of our
future research directions.

Another interesting problem worth to be considered is to determine
if the patterns of daily tweet volumes of political parties (for
example, log-normal distribution) have universal features. If this
is the case, it would be important to determine if  we observe
similar patterns for other events. Indeed, broad distributions of
tweet volume for brand names~\cite{Mathiesen2013} and attentions
to online items~\cite{Miotto2014} have already been reported.
Hence, investigation of dynamics of tweet volumes of various
objects can lead us to check universal features of the dynamics.
Further research will be necessary to determine this point.

Influence of social media on political and social issues is
getting greater and greater. Understanding mathematical nature of
dynamics of collective attention to elections in social media can
enhance our ability to anticipate dynamics of collective attention
to other political or social issues.









%
%
%

\newpage

\begin{table*}[!ht]
\caption{ {\bf Description of Twitter data set.} Time stamps in
\emph{Euro14} and \emph{Italy} are in local time while time stamps
in \emph{Bulgaria13} are in Greenwich Mean Time (GMT). There is a
three-hours difference between GMT and Bulgarian time. $T_i$
represents the initial day of considered data. $T_e$ is the
election day. $T_f$ represents the final day of considered data.
One-day is defined a time interval from 00:00:00 to 23:59:59 in
considered time. $N_T$ represents the total number of considered
tweets for given time interval from $T_i$ to $T_e$-1 posted in
local language. $N_P$ represents the number of considered
political parties.}
\begin{tabular}{|l|l|l|l|l|l|l|l|}
\hline
{\bf Data set} & {\bf $T_i$} & \emph{$T_e$}  & $T_f$  & $N_T$ & {\bf Language} & $N_{P}$ & {\bf Held in} \\
\hline
$Euro14$ & 22 Apr. 2014 & 25 May 2014   & 12 Jun. 2014 & 3,413,214 & Italian & 7 & Italy \\
\hline
$Italy13$ & 1 Jan. 2013 & 23 Feb. 2013  & 3 Mar. 2014 & 3,796,754 & Italian & 6 & Italy \\
\hline
$Bulgaria13$ & 29 Apr. 2013 & 12 May 2013 & 27 May 2013 & 5,817 & Bulgarian & 4 & Bulgaria\\
 \hline
\end{tabular}
\begin{flushleft} 
\end{flushleft}
\label{table1}
\end{table*}

\begin{table*}[!ht]
\caption{ {\bf Description of considered political parties for
each election.} The official sources of election results are
provided on
~\cite{Euro2014Result}(\emph{Euro14}),~\cite{Italy2013Result}(\emph{Italy13}),
and ~\cite{Bulgaria2013Result} (\emph{Bulgaria13}) respectively. }
\begin{tabular}{|l|l|l|l|}
\hline \multicolumn{4}{|l|}{\bf Euro14: European Parliament election 2014, Italy} \\
\hline
{\bf Rank} & {\bf Party} & {\bf Actual votes} & {\bf Leaders} \\
\hline
1 & Partito Democratico (PD) & 11,203,231 & Matteo Renzi  \\
\hline
2 & MoVimento Cinque Stelle (M5S) & 5,807,362 & Beppe Grillo   \\
\hline
3 & Forza Italia (FI) & 4,614,364 & Silvio Berlusconi   \\
\hline
4 & Lega Nord (LN) & 1,688,197 & Matteo Salvini \\
\hline
5 & Nuovo Centrodestra - Unione di Centro (NCD-UdC) & 1,202,350 & Angelino Alfano, Pier Ferdinando Casini  \\
\hline
6 & L'Altra Europa con Tsipras (AET) & 1,108,457 & Alexis Tsipras, Nichi Vendola, Paolo Ferrero \\
\hline
7 & Fratelli d'Italia - Alleanza Nazionale (FdI-AN) & 1,006,513 & Giorgia Meloni \\
 \hline \hline \multicolumn{4}{|l|}{\bf Italy13: Italian general election
 2013}\\
\hline
{\bf Rank} & {\bf Party} & {\bf Actual votes} & {\bf Leaders}  \\
\hline
1 &  MoVimento Cinque Stelle (M5S) & 8,691,406 & Beppe Grillo   \\
\hline
2 & Partito Democratico (PD) & 8,646,034 & Pier Luigi Bersani, Matteo Renzi \\
\hline
3 & Il Popolo della Libert\`{a} (PdL) & 7,332,134 & Silvio Berlusconi  \\
\hline
4 & Scelta Civica (SC) & 2,823,842 & Mario Monti \\
\hline
5 & Lega Nord (LN) & 1,390,534 & Roberto Maroni \\
\hline
6 & Sinistra Ecologia Libert\`{a} (SEL) & 1,089,231 & Nichi Vendola \\
 \hline \hline
\multicolumn{4}{|l|}{\bf Bulgaria13: Bulgarian general
election 2013} \\
\hline
{\bf Rank} & {\bf Party} & {\bf Actual votes} & {\bf Leaders} \\
\hline
1 & GERB & 1,081,605 & Boyko Borisov  \\
\hline
2 & BSP & 942,541 & Sergei Stanishev  \\
\hline
3 & DPS & 400,446 & Lyutvi Mestan \\
\hline
4 & ATAKA & 258,481 & Volen Siderov \\
 \hline
\end{tabular}
\begin{flushleft} 
\end{flushleft}
\label{table2}
\end{table*}

\begin{table*}[!ht]
\caption{ {\bf Parameters to describe the dynamics of daily tweet
volume of political parties as a geometric Brownian motion (GBM).}
The expectation value $V_p(t)$ of daily tweet volume of party $p$
at time $t$ given by a GBM is $V_p(t)=V_p(0)
exp((\mu-\sigma^2/2)t+\sigma W(t))$ where $W(t)$ is a Wiener
process or a Brownian motion.}
\begin{tabular}{|l|l|l|l|l|l|l|l|l|l|l|l|}
\hline \multicolumn{12}{|l|}{\bf Euro14: European Parliament election 2014, Italy} \\
\hline
{\bf Rank} & {\bf Party} & {\bf $\mu-\sigma^2/2$} & {\bf $\mu$} & {\bf $\sigma$} & $V_p(0)$ & {\bf Rank} & {\bf Party} & {\bf $\mu-\sigma^2/2$} & {\bf $\mu$} & {\bf $\sigma$} & $V_p(0)$\\
\hline
1 & PD  & 0.0124 & 0.0627 & 0.3171 & 18299.2 & 5 & NCD-UdC & 0.0059 & 0.0893 & 0.4088 & 2578.5\\
\hline
2 & M5S & 0.0469 & 0.0925 & 0.3018 & 6143.3 & 6 & AET & 0.0581 & 0.1513 & 0.4316 & 520.0\\
\hline
3 & FI  & 0.0053 & 0.0955 & 0.4247 & 9714.3 & 7 & FdI-AN & 0.0404 & 0.4013 & 0.8496 & 238.5  \\
\hline
4 & LN  & 0.0592 & 0.2995 & 0.6932 & 686.2 &   &        &  & & & \\
\hline \hline \multicolumn{12}{|l|}{\bf Italy13: Italian general
election 2013}\\
\hline
{\bf Rank} & {\bf Party} & {\bf $\mu-\sigma^2/2$} & {\bf $\mu$} & {\bf $\sigma$} & $V_p(0)$ & {\bf Rank} & {\bf Party} & {\bf $\mu-\sigma^2/2$} & {\bf $\mu$} & {\bf $\sigma$} & $V_p(0)$\\
\hline
1 & M5S & 0.0328 & 0.0979 & 0.3608 & 3294.9 & 4 & SC & -0.0048 & 0.0490 & 0.3278 & 22856.7\\
\hline
2 & PD  & 0.0181 & 0.0815 & 0.3561 & 9121.2 & 5 & LN & 0.0104 & 0.2264 & 0.6573& 576.0 \\
\hline
3 & PdL & 0.0039 & 0.1164 & 0.4744 & 16763.5 & 6 & SEL & 0.0127 & 0.1406 & 0.5057 & 2458.3  \\
\hline \hline \multicolumn{12}{|l|}{\bf Bulgaria13: Bulgarian
general election 2013} \\
\hline
{\bf Rank} & {\bf Party} & {\bf $\mu-\sigma^2/2$} & {\bf $\mu$} & {\bf $\sigma$} & $V_p(0)$ & {\bf Rank} & {\bf Party} & {\bf $\mu-\sigma^2/2$} & {\bf $\mu$} & {\bf $\sigma$} & $V_p(0)$\\
\hline
1 & GERB & 0.0435 & 0.1591 & 0.4808  & 194.8 & 3 & DPS  & 0.2110 & 0.2496 & 0.2782 & 7.6 \\
\hline
2 & BSP  & 0.0892 & 0.1904 & 0.4498 & 55.8 & 4 & ATAKA & 0.2248 & 0.4020 & 0.5954 & 2.4 \\
\hline
\end{tabular}
\begin{flushleft} 
\end{flushleft}
\label{table3}
\end{table*}

\end{document}